\documentclass{llncs}

\usepackage{ifthen}
\usepackage{color}
\usepackage{amssymb}
\usepackage{tikz}
\usepackage[nameinlink,capitalise]{cleveref}

\usetikzlibrary{positioning,calc}
\usepackage[]{enumitem}
\usepackage[misc]{ifsym}

\newcommand\copyrighttext{%
  \footnotesize \textcircled{c} 11th European Conference on Software Architecture (ECSA 2017), Canterbury, UK \\%
Giaimo, Federico, Berger, Christian, and Kirchner, Crispin. ``Considerations about Continuous Experimentation for Resource-Constrained Platforms in Self-Driving Vehicles.'' \textit{In European Conference on Software Architecture (pp. 84-91)}, Springer, 2017, \url{https://link.springer.com/chapter/10.1007/978-3-319-65831-5\_6}}
\newcommand\copyrightnotice{%
\begin{tikzpicture}[remember picture,overlay]
\node[anchor=south,yshift=10pt] at (current page.south) {\fbox{\parbox{\dimexpr\textwidth-\fboxsep-\fboxrule\relax}{\copyrighttext}}};
\end{tikzpicture}%
}

\begin{document}
\title{Considerations about Continuous Experimentation for Resource-Constrained Platforms in Self-Driving Vehicles}
\titlerunning{}  
\author{Federico Giaimo\inst{1}(\Letter) \and Christian Berger\inst{2}\and Crispin Kirchner\inst{3}}
\authorrunning{Federico Giaimo, Christian Berger, Crispin Kirchner} 
\tocauthor{Federico Giaimo, Christian Berger, Crispin Kirchner}
\institute{Chalmers University of Technology, G\"oteborg, Sweden\\
\email{giaimo@chalmers.se}
\and
University of G\"oteborg, G\"oteborg, Sweden\\
\email{christian.berger@gu.se}
\and
RWTH Aachen University, Germany\\
\email{crispin.kirchner@rwth-aachen.de}
}

\maketitle

\begin{abstract}
Autonomous vehicles are slowly becoming reality thanks to the efforts of many academic and industrial organizations. 
Due to the complexity of the software powering these systems and the dynamicity of the development processes, an architectural solution capable of supporting long-term evolution and maintenance is required.

Continuous Experimentation (CE) is an already increasingly adopted practice in software-intensive web-based software systems to steadily improve them over time.
CE allows organizations to steer the development efforts by basing decisions on data collected about the system in its field of application. 
Despite the advantages of Continuous Experimentation, this practice is only rarely adopted in cyber-physical systems and in the automotive domain.
Reasons for this include the strict safety constraints and the computational capabilities needed from the target systems. 

In this work, a concept for using Continuous Experimentation for resource-constrained platforms like a self-driving vehicle is outlined.


\end{abstract}
\copyrightnotice
\vspace*{-1cm}
\section{Introduction} \label{sec:introduction}

Constant efforts in technology and software development by various research and commercial institutions are making autonomous cars gradually a reality.
While this final objective is still out of reach in the nearest future, many features that can replace the human driver in ordinary driving tasks are already available. 

Due to its safety constraints 
the software in vehicles needs to be very high in quality.
This will prove even more true for autonomous vehicles, which will have the responsibility to assess the real world around them to decide a course of action while always meeting the safety requirements.
For this reason it is imperative to find and enable a process that allows continuous software quality improvements, possibly even after the vehicle is sold to the customers.

Continuous Experimentation (CE) is an Extreme Programming practice that could satisfy these needs by running so-called ``experiments'' to collect meaningful data. 
These experiments are usually either variants of the deployed software or additional software features. 
The goal is to collect and use the resulting real-world data in order to decide in an objective way which of the possible variants or features is the most successful one. 
A CE setup begins with the target-base divided in sets, one of which is the \textit{control set}, running unmodified software, and one or more \textit{experimental sets}, which will run an experiment each. 
The software in all sets 
then collects relevant usage and performance data that will be relayed back to the developers. 
The best-performing set will decide which software variant or feature will be further developed and deployed to all the other targets. 

CE is increasingly adopted in the context of software-intensive web-based applications, and the current state-of-practice is outlined in \cref{sec:related}. 
With a focus on autonomous vehicles, we outlined in our previous work the design criteria for the software architecture to enable experimentation on Cyber-Physical Systems (CPS) as well~\cite{GB17}. 
However, challenges related to safety considerations 
are still unresolved and pose a significant obstacle for the adoption of software experimentation on vehicles. 
Scarcity of resources plays also an important role in this sense since the hardware in the car is carefully dimensioned in terms of performances to provide ``just enough''. 
Further challenges like scalability issues in case of several systems conducting experiments have also been identified in our previous study~\cite{GYBC16}. 

The Research Goal of this work is to assess the challenges related to the scarcity of resources that prevent the widespread adoption of CE in the automotive context, and to propose strategies to overcome them. 

This goal is further elaborated into the following Research Questions: 
\begin{enumerate}[leftmargin=*,labelindent=0em,label=$RQ{\arabic*}:$]
\item What impact does the lack of resources in cyber-physical systems impose on the design and application of continuous experimentation?
\item What design criteria should the software architecture satisfy in order to enable continuous experimentation for a resource-constrained cyber-physical system?
\end{enumerate}


\section{Related Work} \label{sec:related}
Several works are present in literature focusing on Continuous Experimentation. 
One of these is Fagerholm et al.~\cite{FGMM16}, which describes a CE model that takes into account the roles, tasks, infrastructure and information artifacts involved by this practice.
In this paper, the authors developed and extended their model, validating it against the results of two empirical case studies conducted in startup companies. 

Another article of interest is Olsson and Bosch~\cite{OB14}, which describes the steps that should be taken to move a traditional software development process to a ``continuous'' one.
These steps involve the gradual introduction of Agile practices and the modification of the organization and their strategies in order to align them to the ones that better support continuous product evolution and delivery.

Several articles related to CE report the advancements and characteristics of the experimentation processes and platforms in industrial settings.
An example of these works is Tang et al.~\cite{TAOM10} that described the experimental setting at Google~Inc. where, in order to improve the experimentation process and execution, experiments that involve independent factors are overlapped.
Further examples are Kohavi et al.~\cite{KDF+13}, that described Microsoft Bing's own solution to run ``over 200 experiments concurrently'', and Amatriain~\cite{Amatriain13}, that outlined Netflix's approach to experimentation. 

At the best of the authors' knowledge, and perhaps hinting at the novelty of the field, some of the major academic databases, i.e.~IEEE~Xplore, ACM Digital Library, Scopus, Web of Science, were searched for articles regarding Continuous Experimentation in the context of CPS, but unrelated or no results at all were found at the time of writing. 


\section{Assessing the Scarcity of Resources}
\label{sec:scarcity}

Running experimental software alongside production software requires additional computational resources. 
In contrast to web-based applications running in server farms, where additional virtual servers can be spawned if needed, acquiring additional computational power in CPS is not trivial, as their hardware cannot be changed after delivery to the customers.

\begin{figure}
	\centering
	\begin{tikzpicture}[%
            node distance=9mm and .5mm,
            every node/.append style={align=left},
            M/.style={draw,align=left,text width=0.65cm,inner xsep=.5ex,minimum height=2.5em},
            E/.style={fill=black!5,text width=.5cm},
            Delim/.style={inner sep=0},%
        ]
        \newcommand{\RootComponentFamily}{\bfseries}
        \newcommand{\PTimeSliceWidth}{0.95cm}
        \newcommand{\STimeSliceWidth}{1.7125cm}
        \newcommand{\DSTimeSliceWidth}{\PTimeSliceWidth}
        \newcommand*{\Delim}[3]{%
            \node[Delim,xshift=#1] (#3) at(#2.west) {\tikz\draw (0,0) -- (0,2.5em+2mm);}%
        }
        \newcommand*{\PDelim}[2]{%
            \node[Delim,xshift=\PTimeSliceWidth] (PDelim#1) at(PDelimY-|PP#1.west) {\tikz\draw (0,0) -- (0,5em+4mm);};
            \coordinate[right=of PDelim#1] (PTS#2Start);
        }

        \node (DSLabel) {\RootComponentFamily Down-\strut\\\RootComponentFamily sampling\strut};
        \node[node distance=and 0.5em,right=of DSLabel] (DSCPU) {\RootComponentFamily CPU0\strut};
        \coordinate[right=of DSCPU] (TimelineStart);
        \node[M,anchor=north west] (DS0) at(DSLabel.north-|TimelineStart) {{\RootComponentFamily P}\\0};
        \Delim{\DSTimeSliceWidth}{DS0}{DSDelim0};
        \node[M,right=of DSDelim0] (DS1) {{\RootComponentFamily P}\\1};
        \Delim{\DSTimeSliceWidth}{DS1}{DSDelim1};
        \node[M,right=of DSDelim1,E] (DS2) {{\RootComponentFamily E}\\2};
        \Delim{\DSTimeSliceWidth}{DS2}{DSDelim2};
        \node[M,right=of DSDelim2] (DS3) {{\RootComponentFamily P}\\3};
        \Delim{\DSTimeSliceWidth}{DS3}{DSDelim3};
        \node[M,right=of DSDelim3] (DS4) {{\RootComponentFamily P}\\4};
        \Delim{\DSTimeSliceWidth}{DS4}{DSDelim4};
        \node[M,E,right=of DSDelim4] (DS5) {{\RootComponentFamily E}\\5};
        \Delim{\DSTimeSliceWidth}{DS5}{DSDelim5};
        \node[M,right=of DSDelim5] (DS6) {{\RootComponentFamily P}\\6};
        \Delim{\DSTimeSliceWidth}{DS6}{DSDelim6};
        \node[right=of DSDelim6] {$\dots$};

        \node[M,above=of DS0.north west,anchor=south west] (SP0) {{\RootComponentFamily P}\\0};
        \node[anchor=west] (SCPU) at(DSCPU.west|-SP0) {\RootComponentFamily CPU0};
        \node[anchor=north west] (SLabel) at(DSLabel.west|-SP0.north) {\RootComponentFamily Serial};
        \node[M,E,right=of SP0] (SE0) {{\RootComponentFamily E}\\0};
        \Delim{\STimeSliceWidth}{SP0}{SDelim0};

        \node[M,right=of SDelim0] (SP1) {{\RootComponentFamily P}\\1};
        \node[M,E,right=of SP1] (SE1) {{\RootComponentFamily E}\\1};
        \Delim{\STimeSliceWidth}{SP1}{SDelim1};

        \node[M,right=of SDelim1] (SP2) {{\RootComponentFamily P}\\2};
        \node[M,E,right=of SP2] (SE2) {{\RootComponentFamily E}\\2};
        \Delim{\STimeSliceWidth}{SP2}{SDelim2};

        \node[M,right=of SDelim2] (SP3) {{\RootComponentFamily P}\\3};
        \node[M,E,right=of SP3] (SE3) {{\RootComponentFamily E}\\3};
        \Delim{\STimeSliceWidth}{SP3}{SDelim3};

        \node[right=of SDelim3] {$\dots$};

        \node[M,E,above=of SP0.north west,anchor=south west] (PE0) {{\RootComponentFamily E}\\0};
        \node[M,node distance=1mm,above=of PE0.north west,anchor=south west] (PP0) {{\RootComponentFamily P}\\0};
        \coordinate (PDelimY) at($(PP0)!0.5!(PE0)$);
        \node[anchor=west] at(DSCPU.west|-PP0) {\RootComponentFamily CPU0};
        \node[anchor=west] at(DSCPU.west|-PE0) {\RootComponentFamily CPU1};
        \node[anchor=north west] at(DSLabel.west|-PP0.north) (PLabel) {\RootComponentFamily Parallel};
        \PDelim{0}{1};

        \node[M,anchor=west] at(PTS1Start|-PP0) (PP1) {{\RootComponentFamily P}\\1};
        \node[M,anchor=west,E] at(PTS1Start|-PE0) (PE1) {{\RootComponentFamily E}\\1};
        \PDelim{1}{2};

        \node[M,anchor=west] at(PTS2Start|-PP0) (PP2) {{\RootComponentFamily P}\\2};
        \node[M,anchor=west,E] at(PTS2Start|-PE0) (PE2) {{\RootComponentFamily E}\\2};
        \PDelim{2}{3};

        \node[M,anchor=west] at(PTS3Start|-PP0) (PP3) {{\RootComponentFamily P}\\3};
        \node[M,anchor=west,E] at(PTS3Start|-PE0) (PE3) {{\RootComponentFamily E}\\3};
        \PDelim{3}{4};

        \node[M,anchor=west] at(PTS4Start|-PP0) (PP4) {{\RootComponentFamily P}\\4};
        \node[M,anchor=west,E] at(PTS4Start|-PE0) (PE4) {{\RootComponentFamily E}\\4};
        \PDelim{4}{5};

        \node[M,anchor=west] at(PTS5Start|-PP0) (PP5) {{\RootComponentFamily P}\\5};
        \node[M,anchor=west,E] at(PTS5Start|-PE0) (PE5) {{\RootComponentFamily E}\\5};
        \PDelim{5}{6};

        \node[M,anchor=west] at(PTS6Start|-PP0) (PP6) {{\RootComponentFamily P}\\6};
        \node[M,anchor=west,E] at(PTS6Start|-PE0) (PE6) {{\RootComponentFamily E}\\6};
        \PDelim{6}{7};

        \node[anchor=west] (PDots) at(PTS7Start) {$\dots$};

        \coordinate[node distance=1mm,below=of PDelim0] (PTimeAxisY);
        \draw[->] (PTimeAxisY-|PP0.west) -- (PDots.east|-PTimeAxisY) node[anchor=north west] {$t$};

        \coordinate[node distance=1mm,below=of SDelim0] (STimeAxisY);
        \draw[->] (STimeAxisY-|PP0.west) -- (PDots.east|-STimeAxisY) node[anchor=north west] {$t$};

        \coordinate[node distance=1mm,below=of DSDelim0] (DSTimeAxisY);
        \draw[->] (DSTimeAxisY-|PP0.west) -- (PDots.east|-DSTimeAxisY) node[anchor=north west] {$t$};
    \end{tikzpicture}
    
    \caption{Execution strategies. ``P'' and ``E'' stand for Production and Experimental software module. Picture based on Kirchner~\cite{Kirchner17}.}
    \label{fig:ExecutionStrategies}
\end{figure}
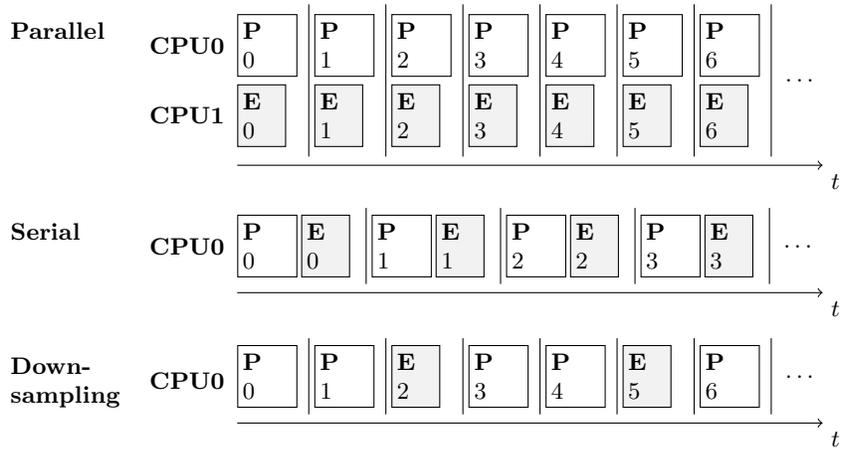

To assess these limitations, different \textit{execution strategies} for acquiring unused computational power are proposed, taking into consideration different initial conditions 
that we have explored in the context of one of our research projects~\cite{Kirchner17}. 
These strategies are explained in the following paragraphs and depicted in \cref{fig:ExecutionStrategies}. 
The automotive software in the proposed execution scenarios is assumed to be structured in modules, which are recurrently executed in time slots, either \textit{data-} or \textit{time-triggered}~\cite{NS08}. 
This means respectively that a module is either executed whenever new information arrives, or at a fixed frequency even if new data has not been gathered or if new data was queued waiting to be processed. 
The ideal way to test an experimental version of a production software module would be to run it in parallel to the production version in order to provide the same input to both modules. 
However, due to safety reasons and lack of computational resources the experimental module may be forced to run on a less frequent schedule than the production module and its communications capabilities may be reduced (for example its output could be logged instead of forwarded to the intended recipients).
In order to make the experimental software ``believe'' that it is being run without such handicaps 
it is required to encapsulate the time and the communication resources that the software modules can access. 

Due to the necessary level of control needed over the software modules 
in the authors' understanding it is not enough to simply delegate the experiment's execution schedule to the operating system's Process Scheduler. 
Firstly because 
the choice of whether to run an experimental module and what execution schedule to adopt depends on several factors that are only known at high levels of abstraction. 
Secondly and more importantly, executing an experiment can imply the execution of a software module at the potential ``expenses'' of another selected one when computational resources are scarce, and to unfairly favor a software module over another is against the Process Scheduler's goal to serve resources in a fair way among all processes. 

In the following the identified execution strategies will be described.
\begin{itemize}[leftmargin=*,labelindent=0em]
\setlength\itemsep{0.3em}
\item[] \textbf{Parallel Execution.}
In the simplest case, even though either time or computational resources are scarce on a particular core or processor alongside the production module, a third software module can be paused or stopped in order to reuse its resources to run the experiment.
In this case it is possible to assume that an unused processor is available, and the experimental module can be executed in \textit{parallel} to the production module. 
As both modules run on independent computing units, they are not necessarily coupled in terms of execution frequency. 
This case has been described for completeness but it is unlikely to be applicable.

\item[] \textbf{Serial Execution.}
In the typical case that there is no additional computing unit available to independently execute an experimental module, the computing time needed by the experiment could come from the unused time of a production module. 
In this case the experimental module could be executed \textit{serially}, i.e.~always after the production module has finished its computation and until the production module is executed again in its next time slot. 

When production and experimental modules are functionally related and are supposed to operate as synchronously as possible, two different cases with different implications can be identified: whether the experimental module can or cannot conclude its calculations in the unused time left in the production module's time slot.
In the simplest case, the experimental module can finish its tasks inside the time window left over by the production module, 
in the second case, the time left unused by the production module is not enough for the experimental module to complete its operations, which results in an interruption of the experimental module. 
It is worth noting that whenever the execution of the experimental module needs to be stretched over two non-contiguous time slots due to the lack of unused time in the current slot, the result is that the experimental module will be executed less frequently than the production module, potentially resulting in time synchronization issues and affecting the comparability of metrics in the case of A/B testing.

\item[] \textbf{Downsampled Execution.}
The third execution strategy, called \textit{downsampling}, is applicable if there is no additional computing node available and no computation time is left in the time slice of a module. 
As computational power on cars is limited, it can be expected to also be the most likely applicable strategy. 
This approach is based on the assumption that conditions exist under which the execution of a production module can periodically be skipped (analog to suspending the production module from time to time), freeing computational resources to be used for experimentation purposes. 
Skipping execution cycles of a production module may result in compromising safety-critical aspects of the vehicle, hence great care must be taken to ensure that the planned downsampling is safe. 
A possible way to ensure its safety could be to run preliminary tests before applying this strategy, to verify in advance that it is viable in practice and at which rate the production module can skip computation cycles before dependent modules downstream in the data-processing chain are affected. 
Furthermore, the conditions under which the downsampling rate has been tested need to be fixed and the execution of the experiment must only be carried out when the vehicle operates under those conditions. 
As with this strategy the time slots available to the experimental module are non-contiguous, the considerations about time synchronization and logic coherence that were expressed for the serial execution strategy apply to this case as well. 
\end{itemize}

The proposed strategies may also be composed and adjusted at runtime. 
For example, it could happen that an experiment might initially require the analysis of relatively small amounts of data, thus making the serial execution strategy feasible.
If however more intensive calculations would later be required and the conditions would allow it, the strategy could be changed to downsampling in order to allocate more time to each experimental iteration at the cost of a less frequent execution schedule.


\section{Software Architecture} 
\label{sec:software}
\cref{sec:scarcity} has identified three potential strategies to execute an experimental software module next to a product module. 
Furthermore, we have pointed out that the production and experimental modules need to be decoupled from the real system time and from their respective potential communication vector with downstream modules.
The reason is that 
the production and experimental module should believe that they are triggered at the very same point in time by the same input data; 
while the execution strategy in effect must be entirely transparent for the modules.
Also, the communication of data into and from the production and experimental modules 
must be controlled entirely. 
While the ingoing communication may not be 
critical, 
a strict control of any outgoing communication is needed to avoid unwanted interference with the dependent downstream software modules. 
Furthermore, any time stamping related to sending data from the production and experimental modules to other modules must be potentially adjusted to make the rest of the system believe that these modules have not been executed with different execution strategies. 
The possibility of \textit{rewriting} time stamp information for communication is another indicator why the regular Process Scheduler provided by the operating system does not meet the requirements for conducting experiments on a resource-constrained computational environment.

Chalmers University of Technology hosts a vehicle laboratory called Revere, ``Resource for Vehicle Research''~\cite{URL_revere}, with the goal of conducting and developing research for self-driving vehicles and active safety. 
The Revere laboratory uses our middleware OpenDaVINCI\footnote{\url{http://code.opendavinci.org}}, which allows the realization of distributed microservices communicating via Protobuf-encoded messages. 
The activation of software modules realized with OpenDaVINCI complies to the time-triggered
or data-triggered 
principle described in \cref{sec:scarcity}.
OpenDaVINCI by default encapsulates the system time via an object called \texttt{TimeStamp} 
that either invokes the POSIX time API 
returning the ``real'' time or transparently replaces the real system clock with a virtual one. 
The communication facilities available to the software modules are also encapsulated. 
OpenDaVINCI uses by default UDP multicast as communication principle. 
In OpenDaVINCI a so-called \texttt{ContainerConference} is provided as the data to be exchanged is wrapped into \texttt{Container} containing the actual data to be exchanged and some meta-information like time stamps for sending, receiving, and sample time point. 

To enable CE using these building blocks, both the production and experimental modules 
will be handled by an \textit{Experimenter} software module that will manage them 
to realize the aforementioned execution strategies by forwarding input data to both modules, activating and suspending them according to the respective execution strategy, and receiving data containers to be distributed for both delivery or logging purposes.


\section{Discussion} 
\label{sec:discussion}
For the current state-of-practice of CE in web-based systems, which usually involves validation of user feedback, small scale approaches are not viable 
since less generalizable.
However, in the automotive domain the experiments would 
focus on algorithmic problems and their verification in realistic scenarios, making the results 
easier to generalize even if collected by a small number of vehicles. 

This work proposes a new element to consider in order to apply CE on cyber-physical systems, which is the execution strategy. 
This element is introduced to account for the possible lack of computational resources, and can critically impact the amount of collected results or the overall viability of the experiments. 
For this reason we propose an addition to the CE model proposed by Fagerholm et al.~\cite{FGMM16} when it involves experiments on CPS: the \textit{domain expert}, a person or team with deep knowledge of the system and its capabilities.
The domain expert's main role is to advise the experimenter and data scientist while devising and planning the experiment to be run. 
The insights this figure could provide would not be limited only to the choice of the execution strategy but could range for example from deciding if an experiment could be run ``live'' on customers' vehicles, or if preliminary measurements would be needed to ensure its viability, and so on.
As a direct application of the ``web-based'' continuous experimentation would prove difficult or even impossible in the context of CPS due to the several key differences between the two fields,
we claim that the presence of an intermediary figure can smoothen or in some cases enable the experimentation process thanks to its knowledge of both the system and the proposed techniques to obtain the additional computational time needed to run experiments.

We report about threats to the validity of this study according to Runeson and H\"{o}st \cite{RH09}. 
Our current work in the lab concerns the validation of the proposed strategies using our self-driving vehicles to increase the external validity of the suggested architectural design considerations.
It is also impossible to completely eliminate the threat to reliability, i.e.~whether different researchers would come out with the same solution if they were to assess the same problem. 
To mitigate this threat, we carefully described our reasoning to motivate our suggested design decisions.


\section{Conclusions and Future Work} 
\label{sec:conclusions}

The present work aims at contextualizing the Continuous Experimentation process into the Cyber-Physical System field, assessing the lack of surplus resources that would be needed for the system to run the additional experimental code.
In order to assess this deficit, three different execution strategies have been proposed that would allow to run an experimental software module alongside a production module. 
The different characteristics of the strategies enable the adaptation of the solution for different application scenarios. 

In order for a software architecture to enable and make use of the proposed strategies it must be possible to strictly control two crucial types of information that are accessible to both the production and experimental software module, which are the time and the communication resource. 
Controlling the modules' access to these resources acts as enabling criteria ensuring the transparency of the execution strategy to the software modules themselves. 

Future efforts will focus on evaluating the contributions in a setting closer to the specific challenges encountered in industry, by continuing the research in the COPPLAR project, which is Chalmers University of Technology's contribution to the DriveMe context\footnote{\url{http://www.chalmers.se/en/areas-of-advance/Transport/news/Pages/Chalmers-joins-the-Drive-Me-project.aspx}}.
The DriveMe project is an autonomous driving pilot project by Volvo Cars that aims at releasing 100 cars capable of self-driving capabilities on selected public roads in 2017.

\section*{Acknowledgment}
This work has been supported by the COPPLAR Project -- CampusShuttle cooperative perception and planning platform~\cite{URL_copplar}, funded by Vinnova FFI, Diarienr: 2015-04849.


\bibliographystyle{splncs}
\bibliography{bib/library}

\end{document}